\documentclass[lettersize,journal]{IEEEtran}
\usepackage{amsmath,amsfonts}
\usepackage{esint}
\usepackage{algorithmic}
\usepackage{algorithm}
\usepackage{array}
\usepackage[caption=false,font=normalsize,labelfont=sf,textfont=sf]{subfig}
\usepackage{textcomp}
\usepackage{stfloats}
\usepackage{url}
\usepackage{verbatim}
\usepackage{graphicx}
\usepackage{cite}
\begin{document}

\title{Spontaneous emission rate and the density of states inside a one dimensional photonic crystal}

\author{Ebrahim Forati,~\IEEEmembership{Member,~IEEE}
\thanks{This work is done as an independent researcher (email: forati@ieee.org.)}}

\markboth{}%
{Shell \MakeLowercase{\textit{et al.}}: A Sample Article Using IEEEtran.cls for IEEE Journals}


\maketitle

\begin{abstract}
Different densities of electromagnetic states inside a one dimensional
photonic crystal (1D PC) are studied. Hertz vector formalism is used
to calculate Green's tensor inside a layered structure, semi-analytically.
Based on the obtained Green's tensor, the local density of electromagnetic
states (LDOS) and the density of states (DOS) inside a 1D PC are calculated
and discussed. The Green's tensor is also used to approximate the
density of Bloch states inside the 1D PC and is compared with its
exact calculation based on the 1D PC dispersion relations. Using a
practical 1D PC parameters in the visible range, the aforementioned
quantities are calculated and verified with a full-wave solver based
on finite element method (FEM). The formulations and the results are
aimed to be helpful in thermal and spontaneous radiation studies. 
\end{abstract}

\begin{IEEEkeywords}
Density of electromagnetic states, photonic crystal, Green's function.
\end{IEEEkeywords}

\section{Introduction}

\IEEEPARstart{T}{he}  density of electromagnetic modes/states (DOS) in a homogeneous
isotropic medium is defined as the number of available electromagnetic
modes per volume per frequency. It is an important quantity in statistical
physics including thermal energy radiation where the degeneracy of
the energy levels (related to the photons' frequencies) is a determining
parameter in the thermodynamical arguments. The DOS is often used
as the degeneracy function in such thermal radiation studies. A well-known
example is the conventional broadband thermal radiation which obeys
Planck\textquoteright s blackbody radiation \cite{kirchhoff1860ueber,planck2013theory}
(i.e. radiations from Sun, a tungsten light bulb, etc.) The DOS also
has significant importance in quantum electrodynamics (QED) where
physics of spontaneous emission, vacuum fluctuations, Van der Waals
and Casimir forces, etc. depend on the environment DOS. Note, the
DOS in this work refers to the photonic states. The term is also used
for the density of available electronic states in solid state physics. 

A closely related parameter to the DOS is the local density of electromagnetic
states (LDOS) which is defined to be proportional to the energy released
by a unity dipole transition (point current source) per second. The
released energy travels away from the source as photons. For a point
current source at $\mathbf{r_{0}}$ , radial frequency of $\omega,$
and in the direction of $\hat{\alpha}$, $\mathbf{J}\left(\mathbf{r^{\prime}},\omega^{\prime}\right)=\widehat{\mathbf{\alpha}}\delta\left(\mathbf{r^{\prime}}-\mathbf{r}_{0}\right)\delta\left(\omega^{\prime}-\mathbf{\omega}\right),$
the LDOS is 

\begin{equation}
\mathrm{LDOS}\left(\mathbf{r_{0}},\omega,\widehat{\mathbf{\alpha}}\right)=\frac{12\varepsilon_{0}\varepsilon_{r}}{\pi}\mathrm{P}\left(\mathbf{r_{0}},\omega,\widehat{\mathbf{\alpha}}\right)\label{eq:LDOS_1}
\end{equation}
where $\mathrm{P}\left(\mathbf{r_{0}},\omega,\widehat{\mathbf{\alpha}}\right)=\frac{1}{2}\operatorname{Re}\left\{ \mathbf{E}\left(\mathbf{r_{0}},\omega\right)\cdot\mathbf{J}\left(\mathbf{r_{0}},\omega\right)\right\} $
is the power released by $\mathbf{J}\left(\mathbf{r_{0}},\omega\right)$,
and $\varepsilon_{r}$ is the relative permittivity of the medium
($\varepsilon_{0}$ and $\mu_{0}$ are the free space permittivity
and permeability, respectively.) It is common to express (\ref{eq:LDOS_1})
in terms of the dyadic Green's tensor as \cite{economou2006green}

\begin{equation}
\mathrm{LDOS}\left(\mathbf{r_{0}},\omega,\widehat{\mathbf{\alpha}}\right)=\frac{6\omega\mu_{r}\varepsilon_{r}}{\pi c^{2}}\left|\operatorname{Im}\left(\hat{\alpha}\cdot\overline{\overline{\mathbf{G}}}\left(\mathbf{r}_{0},\mathbf{r_{0}},\omega\right)\cdot\hat{\alpha}\right)\right|,\label{eq:LDOS2}
\end{equation}
where the Green's tensor, in an isotropic medium, satisfies 

\begin{equation}
\nabla\times\nabla\times\overline{\overline{\mathbf{G}}}\left(\mathbf{r}_{0},\mathbf{\mathbf{r}^{\prime}},\omega\right)-k_{0}^{2}\mu_{r}\varepsilon_{r}\mathbf{\overline{\overline{\mathbf{G}}}\left(\mathbf{r}_{0},\mathbf{\mathbf{r}^{\prime}},\omega\right)}=\delta\left(\mathbf{r_{0}}-\mathbf{r}^{\prime}\right)\mathbf{\overline{\overline{\mathbf{I}}},}\label{eq:GF}
\end{equation}
in which $k_{0}=\omega\sqrt{\varepsilon_{0}\mu_{0}}$ is the free
space wavenumber, $\overline{\overline{\mathbf{I}}}$ is the dyadic
unity matrix, $\mu_{r}$ is the relative permeability of the medium
at the source location, and $c$ is the speed of light in free space.
We use $e^{j\omega t}$ time harmonic convention throughout the paper,
and the primed notation is used for the source parameters. The electric
field is related to the Green's tensor as 

\begin{equation}
\mathbf{E}\left(\mathbf{r},\omega\right)=j\omega\mu_{0}\mu_{r}\intop_{\mathbf{r^{\prime}}}\overline{\overline{\mathbf{G}}}\left(\mathbf{r},\mathbf{\mathbf{r^{\prime}}},\omega\right)\cdot\mathrm{\mathbf{J}}\left(\mathbf{r^{\prime}},\omega\right)d\mathbf{r^{\prime}}.\label{eq:electric_greens}
\end{equation}
It is straight forward to convert (\ref{eq:LDOS_1}) to (\ref{eq:LDOS2})
since $\hat{\alpha}\cdot\mathbf{E}\left(\mathbf{r_{0}},\omega\right)=j\omega\mu_{0}\mu_{r}\left(\hat{\alpha}\cdot\overline{\overline{\mathbf{G}}}\left(\mathbf{r}_{0},\mathbf{r_{0}},\omega\right)\cdot\hat{\alpha}\right).$

Note, we also assumed that, for our purpose, the medium's responses
to an electric and a magnetic point sources are similar, hence the
LDOS can be calculated by only considering an electric point source.
In general, (\ref{eq:LDOS2}) should be written as \cite{joulain2003definition}
\begin{equation}\label{eq:LDOS}
\begin{aligned}
\mathrm{LDOS}\left(\mathbf{r_{0}},\omega,\widehat{\mathbf{\alpha}}\right) & =\frac{3\omega\mu_{r}\varepsilon_{r}}{\pi c^{2}}\left(\right.\left|\operatorname{Im}\left(\hat{\alpha}\cdot\mathbf{\overline{\overline{\mathbf{G}}}_{E}}\left(\mathbf{r}_{0},\mathbf{r_{0}},\omega\right)\cdot\hat{\alpha}\right)\right| \\
&\quad + \left|\operatorname{Im}\left(\hat{\alpha}\cdot\mathbf{\overline{\overline{\mathbf{G}}}_{H}}\left(\mathbf{r}_{0},\mathbf{r_{0}},\omega\right)\cdot\hat{\alpha}\right)\right|\left)\right.
\end{aligned}
\end{equation}
where both $\mathbf{\overline{\overline{\mathbf{G}}}_{E}}\left(\mathbf{r}_{0},\mathbf{r_{0}},\omega\right)$
and $\mathbf{\overline{\overline{\mathbf{G}}}_{H}}\left(\mathbf{r}_{0},\mathbf{r_{0}},\omega\right)$
satisfy (\ref{eq:GF}), and $\hat{\alpha}\cdot\mathbf{H}\left(\mathbf{r_{0}},\omega\right)=j\omega\varepsilon_{0}\varepsilon_{r}\left(\hat{\alpha}\cdot\mathbf{\overline{\overline{\mathbf{G}}}_{H}}\left(\mathbf{r}_{0},\mathbf{r_{0}},\omega\right)\cdot\hat{\alpha}\right)$.
To keep the formulations simpler, we continue with (\ref{eq:GF})
throughout the rest of this paper. The effect of this approximation
on the final results will be discussed later. 

LDOS is a critical parameter in nano-photonics and quantum optics
\cite{lodahl2015interfacing,li2018nanophotonic,hsieh2015experimental}.
In particular, the spontaneous emission rate of an emitter in a medium
is proportional to the LDOS via Fermi's golden rule. As a known example,
the spontaneous emission prohibition in photonics crystals (PC) is
achieved by engineering LDOS=0 around the transition frequency of
the emitters \cite{englund2005controlling,joannopoulos1997photonic,kleppner1981inhibited,sanchez2005spontaneous,yablonovitch1994photonic}.
Unlike DOS, the LDOS has both location and direction dependences,
as well as its implicit dependence on the density of available states
and their couplings to the source. In a homogeneous isotropic medium,
DOS and LDOS are essentially the same. 

Although DOS can only be defined, accurately, in a homogeneous isotropic
material, we may estimate it in a periodic structure using the LDOS
as \cite{hasan2018finite}

\begin{equation}
\mathrm{DOS}\left(\omega\right)=\frac{1}{V}\intop_{V}\frac{1}{3}\operatorname{Tr}\left(\mathrm{LDOS}\left(\mathbf{r},\omega,\widehat{\mathbf{\alpha}}\right)\right)d\mathbf{r},\label{eq:DOS-1}
\end{equation}
where $V$ is the volume of the unit cell in the periodic structure,
and the $\frac{1}{3}\operatorname{Tr}\left(.\right)$ operator takes
the average on the three spatial orthonormal directions. 

The estimated DOS of a photonic crystal (PC) is of particular interest
as its thermal emission is argued to surpass Planck's limit \cite{hsieh2015experimental,chow2006theory,cornelius1999modification}.
In fact, there are experimental evidences of emission beyond Planck's
limit by PCs (as high as 100 times) around their band edges \cite{thompson2018hundred}.
One of the most recent experimental confirmations is reported in \cite{lin2020situ}
where emission from a tungsten PC and a blackbody sample are compared
under the same conditions. They observed sharp emission peaks from
the PC (relative to the blackbody), similar to those in DOS of an
unbounded PC . However, finding DOS inside a semi-infinite PC (emitting
into the free half-space above it) needs a careful treatment, as the
electromagnetic modes of the two half-spaces are different and only
partially coupled. In fact, there are wave-guided modes inside a 1D
PC which cannot couple to the outside environment. Examples of such
analysis are \cite{john2008metallic,chow2006theory,luo2004thermal}
only some of which support the possibility of thermal emission beyond
Planck's limit. 

The accurate calculation of the LDOS including its polarization dependence
(and its consequent DOS) inside a 1D PC is a challenging mathematical
problem. Several studies have considered the on-axis modes (plane
waves traveling perpendicular to the 1D PC boundaries) densities,
which have band-gaps and lead to sharp peaks in the calculated LDOS
\cite{boedecker2003all,yeganegi2014local}. However, the total LDOS
of a 1D PC does not have any band-gap \cite{joannopoulos1995photonic}
and the term LDOS in such works should be inferred as the local density
of on-axis states. The LDOS calculations in \cite{prosentsov2007local}
also uses Green's functions with some approximations (called Local
Perturbation Method) and ignores the source's polarization (Green's
function instead of Green's tensor). In \cite{wubs2002local,wubs2004spontaneous},
dielectric slabs are modeled with infinitely thin plane scatterers,
and a multiple scattering formalism is set up to calculate the LDOS.
The polarization dependence of the LDOS is preserved in this formulation,
but a simplified model (i.e. an array of infinitely thin plane scatterers)
is used instead of the 1D PC.

In \cite{brueck2003radiation}, Hertz vector formalism is used to
formulate the exact Green's tensor inside a layered structure (the
exact formulation is also derived for a single slab in \cite{brueck2000radiation}).
The integrations required in the Green's tensor calculation (for inverse
Fourier transforms) are also solved analytically using the complex
plane analysis (as Sommerfeld did, see \cite{sommerfeld1949partial,moore1961dipole})
and mapping techniques such as finding the steepest descent path.
This calculation is exact, but it includes reflection coefficients
from the stacks above and below the source layer (viz. the layer hosting
the point source) as well as the transmission coefficients from the
source layer to the outermost layers. This is great for when the reflection/transmission
coefficients calculations are convenient (e.g. thin film stacks, mirrors,
etc.) However, finding the reflection/transmission coefficients is
a separate problem and is not necessarily an easy task. The same argument
applies to \cite{hoekstra2005theory}. 

In this paper, Hertz vector formalism is used to formulate Green's
tensor inside a layered structure in a matrix form, which is then
solved numerically to find Hertz vectors in all the layers. The advantage
of this formulation is that no reflection/transmission coefficients
are needed to be calculated separately. Also, the integrations are
solved numerically by adding a small loss to the structure (to avoid
singularities). The formulation is intended to be more convenient
for scripting and to be easily scalable for High Performance Computing
(HPC). 

In the followings, first the exact dyadic Green's tensor of a layered
structure are formulated. Next, LDOS and DOS inside a 1D PC are calculated.
We will also discuss the calculation of the density of Bloch modes
using our layered media approach, as well as its exact calculation
based on the 1D PC dispersion relations. Using a practical 1D PC example,
these different mode densities are calculated and compared. The results
are also compared with the numerical calculations based on finite
element method (FEM). The implications of the results to thermal emission
by a 1D PC will also be discussed briefly. 

\section*{LDOS and DOS inside a 1D PC }

\begin{figure}
\centering
\includegraphics[width=5.5cm]{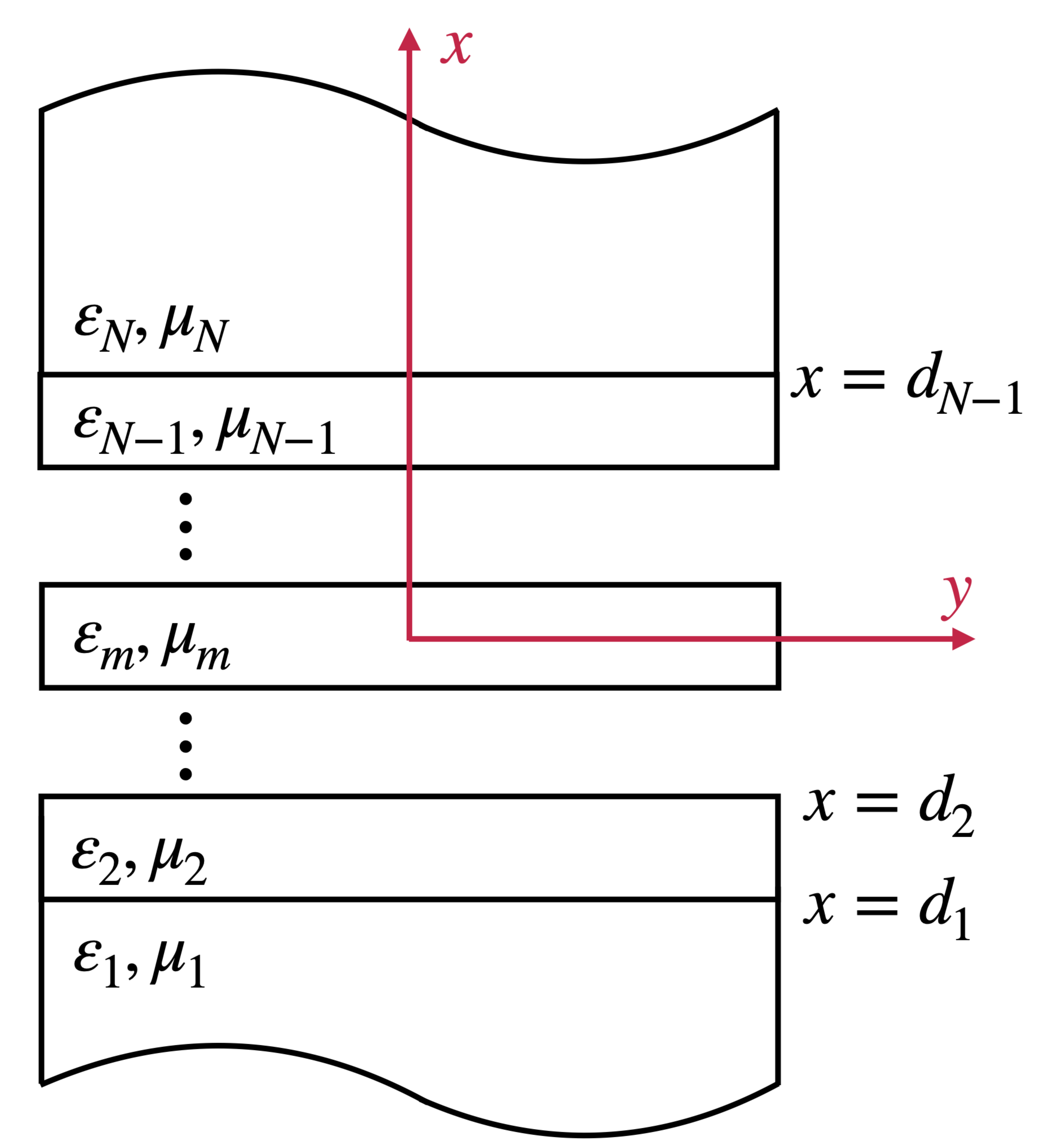}~~\caption{\label{fig:geometry}A one dimensional layered structure. The point
source is at the origin inside layer m (i.e. the coordinate system
moves according to the source location). Layers 1 and N in x-direction,
and all layers in y- and z- directions, are infinitely large. $\varepsilon$
and $\mu$ are relative permittivity and permeability of the layers,
respectively.}
\end{figure}

We start by the Green's tensor expression (using Fourier transform)
inside a layered structure shown in Fig. \ref{fig:geometry}

The first and last layers are infinitely large in x- direction, the
source is at the origin inside the mth layer, and the boundaries are
at $x=d_{1,2,...,\left(N-1\right)}$. Defining two dimensional spatial
Fourier transform pair as 

\begin{equation}
f\left(x,y,z\right)=\frac{1}{\left(2\pi\right)^{2}}\iintop_{-\infty}^{\infty}F\left(x,q_{y},q_{z}\right)e^{j\left(q_{y}y+q_{z}z\right)}dq_{y}dq_{z},\label{eq:Fourier}
\end{equation}

\begin{equation}
F\left(x,q_{y},q_{z}\right)=\iintop_{-\infty}^{\infty}f\left(x,y,z\right)e^{-j\left(q_{y}y+q_{z}z\right)}dydz,
\end{equation}
we may write the nth layer's wave equation in terms of the Hertzian
dipole, $\mathbf{\mathbf{\pi_{n}}}=\mathbf{\mathrm{\mathbf{\mathrm{\pi}_{\mathrm{n}}^{\mathrm{x}}}}}\hat{x}+\mathbf{\mathbf{\mathrm{\pi_{n}^{y}}}}\hat{y}+\mathrm{\mathbf{\mathrm{\mathbf{\mathrm{\mathrm{\pi}}_{\mathrm{n}}^{\mathrm{z}}}}}}\hat{z},$
in Fourier transform domain as \cite{hanson2013operator,hanson2003green}

\begin{equation}
\begin{aligned}
\left(\frac{\partial^{2}}{\partial x^{2}}-p_{n}^{2}\right)\mathbf{\mathbf{\pi_{n}}}\left(x,q_{y},q_{z}\right) & = \\
-\frac{1}{j\omega\varepsilon_{0}\varepsilon_{n}} & \mathbf{J}\left(x,q_{y},q_{z}\right)\delta^{Kron}\left(n-m\right)\label{eq:Hertzian dipole}
\end{aligned}
\end{equation}
where $\delta^{Kron}\left(.\right)$ is the Kronecker delta function,
and $p_{n}=\sqrt{q_{y}^{2}+q_{z}^{2}-k_{0}^{2}\mu_{n}\varepsilon_{n}},$ and the layer m contains the source. 
Without loss of generality, we assume the current is zero in z- direction
($\mathbf{J}=J^{x}\hat{x}+J^{y}\hat{y}$), which also turns out to
give $\pi_{n}^{z}=0.$ The solution to (\ref{eq:Hertzian dipole})
is \cite{hanson2003green,hanson2003leaky,hanson2013operator,yakovlev2003fundamental,bagby1987dyadic}

\begin{equation}
\begin{aligned}
\pi_{n}^{i}\left(x,q_{y},q_{z}\right) & =A_{n}^{i}e^{-p_{n}x}+B_{n}^{i}e^{p_{n}x} \\
& +\delta^{Kron}\left(n-m\right)\intop_{-\infty}^{\infty}\frac{e^{-p_{n}\left|x-x^{\prime}\right|}}{2p_{n}}\frac{J^{i}\left(x^{\prime}\right)}{j\omega\varepsilon_{0}\varepsilon_{n}}dx^{\prime},
\end{aligned}
\end{equation}
where $i=x,y$, and $A_{n}^{i},$ $B_{n}^{i}$ are the unknown coefficients
to be determined from the boundary conditions. 
\par

The electric and magnetic fields are related to the Hertzian dipoles,
in the matrix form, as

\begin{equation}
\begin{aligned}
&\mathbf{E}_{n}\left(x,q_{y},q_{z}\right)=\\
&\left[\begin{array}{ccc} \hat{x} & \hat{y} & \hat{z}\end{array}\right]\left[\begin{array}{ccc}
k_{0}^{2}\mu_{n}\varepsilon_{n}+\frac{\partial^{2}}{\partial x^{2}} & jq_{y}\frac{\partial}{\partial x}\\
jq_{y}\frac{\partial}{\partial x} & k_{0}^{2}\mu_{n}\varepsilon_{n}-q_{y}^{2}\\
jq_{z}\frac{\partial}{\partial x} & -q_{y}q_{z}
\end{array}\right]\left[\begin{array}{c}
\pi_{n}^{x}\\
\pi_{n}^{y}
\end{array}\right],\label{eq:E_pi}
\end{aligned}
\end{equation}
\par

\begin{equation}
\begin{aligned}
&\mathbf{H}_{n}\left(x,q_{y},q_{z}\right)=\\
&j\omega\varepsilon_{0}\varepsilon_{n}\left[\begin{array}{ccc}
\hat{x} & \hat{y} & \hat{z}\end{array}\right]\left[\begin{array}{ccc}
0 & -jq_{z}\\
jq_{z} & 0\\
-jq_{y} & \frac{\partial}{\partial x}
\end{array}\right]\left[\begin{array}{c}
\pi_{n}^{x}\\
\pi_{n}^{y}
\end{array}\right].\label{eq:H}
\end{aligned}
\end{equation}

The boundary conditions are the continuity of tangential fields (electric
and magnetic) at $x=d_{1,2,...,\left(N-1\right)}$, as well as $A_{1}^{x,y}=B_{N}^{x,y}=0$.
This provides a system of $4\left(N-1\right)$ equations and the same
number of unknowns, which can be solved numerically. 

Using (\ref{eq:electric_greens}), (\ref{eq:Fourier}), and (\ref{eq:E_pi}),
the required components of the dyadic Green's tensor for the LDOS
calculations are

{\small{}
\begin{equation}
\begin{aligned}
&\hat{x}\cdot\overline{\overline{\mathbf{G}}}\left(\mathbf{r_{0},r_{0},}\omega\right)\cdot\hat{x}=\\
& \left(\frac{1}{2\pi}\right)^{2}\iintop_{-\infty}^{\infty}dq_{y}dq_{z}\frac{\left(k_{0}^{2}\mu_{m}\varepsilon_{m}+p_{m}^{2}\right)}{j\omega\mu_{0}\mu_{m}}\left(\frac{1}{2p_{m}j\omega\varepsilon_{0}\varepsilon_{m}}+A_{m}^{x}+B_{m}^{x}\right),\label{eq:Gxx}
\end{aligned}
\end{equation}
}{\small\par}

{\small{}
\begin{equation}
\begin{split}
& \hat{y}\cdot\overline{\overline{\mathbf{G}}}\left(\mathbf{r_{0},r_{0},}\omega\right)\cdot\hat{y}  =\left(\frac{1}{2\pi}\right)^{2}\iintop_{-\infty}^{\infty}dq_{y}dq_{z} \Bigg\{ \left(B_{m}^{x}-A_{m}^{x}\right)\frac{q_{y}p_{m}}{\omega\mu_{0}\mu_{m}} \\
&+
 \frac{\left(k_{0}^{2}\mu_{m}\varepsilon_{m}-q_{y}^{2}\right)}{j\omega\mu_{0}\mu_{m}}\left(\frac{1}{2j\omega\varepsilon_{0}\varepsilon_{m}p_{m}}+B_{m}^{y}+A_{m}^{y}\right) \Bigg\} .\label{eq:Gyy}
\end{split}
\end{equation}
}{\small\par}

Note that we had chosen the source to be at $y=z=0$, therefore $e^{j\left(q_{y}y+q_{z}z\right)}=1.$
It will also be helpful to use the replacement $\left(q_{y},q_{z}\right)=\left(q_{\Vert}cos\theta,q_{\Vert}sin\theta\right)$
in (\ref{eq:Gxx}) and (\ref{eq:Gyy}), 

\begin{equation}
\begin{aligned}
\hat{x}\cdot\overline{\overline{\mathbf{G}}}\left(\mathbf{r_{0},r_{0},}\omega\right)\cdot\hat{x}&=\frac{1}{2\pi}\intop_{0}^{\infty}dq_{\Vert}\\
&\frac{q_{\Vert}^{3}}{j\omega\mu_{0}\mu_{m}}\left(\frac{1}{2j\omega\varepsilon_{0}\varepsilon_{m}p_{m}}+A_{m}^{x}+B_{m}^{x}\right),\label{eq:Gxx_simplified}
\end{aligned}
\end{equation}

\begin{equation}
\begin{split}
&\hat{y}\cdot\overline{\overline{\mathbf{G}}}\left(\mathbf{r_{0},r_{0},}\omega\right)\cdot\hat{y}=\left(\frac{1}{2\pi}\right)^{2}\intop_{0}^{\infty}\intop_{0}^{2\pi}d\theta dq_{\Vert}\\
& \Bigg\{ \frac{q_{\Vert}\left(k_{0}^{2}\mu_{m}\varepsilon_{m}-q_{\Vert}^{2}cos^{2}\theta\right)}{j\omega\mu_{0}\mu_{m}}\left(\frac{1}{2j\omega\varepsilon_{0}\varepsilon_{m}p_{m}}+B_{m}^{y}+A_{m}^{y}\right)\\
&+ \left(B_{m}^{x}-A_{m}^{x}\right)\frac{q_{\Vert}^{2}cos\theta p_{m}}{\omega\mu_{0}\mu_{m}}
\Bigg\} .\label{eq:Gyy-2}
\end{split}
\end{equation}
Note that $A_{m}^{x}$ and $B_{m}^{x}$ in (\ref{eq:Gxx_simplified})
are independent of $\theta$ due to the symmetry of the geometry for
a vertical (x- directed) source. This simplifies (\ref{eq:Gxx_simplified})
to only include a one-fold integration. Alternatively, we could obtain
similar results for the vertical source by using Sommerfeld's integral
form of Green's function ($i.e.\intop e^{-p\left|x-x^{\prime}\right|}\frac{J_{0}\left(\rho q_{\Vert}\right)}{2p}dq_{\Vert}$).
However, the spectral form of Green's function, used here, is more
appropriate for a general problem with a randomly oriented source. 

The integrands of (\ref{eq:Gxx_simplified}) and (\ref{eq:Gyy-2})
are the wave components departing the source with the horizontal (parallel
to the boundary) wave-number of $q_{\Vert}$. A wave component becomes
attenuating (as it travels away from the source) if $q_{\Vert}>k_{0}\sqrt{\mu_{m}\varepsilon_{m}}$.
Such a wave component cannot carry energy away from the source unless,
before its complete attenuation, it reaches a region with a supported
$q_{\Vert}$ mode (non-attenuating). Examples are when the point source
is in the vicinity of an interface supporting a surface wave, or near
a dielectric slab supporting a wave-guided mode. 

If the dielectrics in the layered structure are lossless, i.e. $\varepsilon_{n}$
and $\mu_{n}$ are real numbers, the only supported modes by the structure
are Bloch modes and the wave-guided modes (confined within the layers
surrounded by lower permittivity layers.) Such modes have wave-numbers
smaller than $k_{0}\sqrt{\mu\varepsilon_{>}}$ where $\mu\varepsilon_{>}$
is the maximum value of $\mu_{n}\varepsilon_{n}$ among the layers.
This means the integration limits in (\ref{eq:Gxx_simplified}) and
(\ref{eq:Gyy-2}) can be truncated to $k_{0}\sqrt{\mu\varepsilon_{>}}$, 

\begin{equation}
\begin{aligned}
&\operatorname{Im}\left\{ \hat{x}\cdot\mathbf{\overline{\overline{\mathbf{G}}}}\left(\mathbf{r_{0},r_{0},}\omega\right)\cdot\hat{x}\right\}=\\
&\intop_{0}^{k_{0}\sqrt{\mu\varepsilon_{>}}}dq_{\Vert}Re\left\{ \frac{-1}{2\pi}\frac{q_{\Vert}^{3}}{\omega\mu_{0}\mu_{m}}\left(\frac{1}{2j\omega\varepsilon_{0}\varepsilon_{m}p_{m}}+A_{m}^{x}+B_{m}^{x}\right)\right\} ,\label{eq:Gxx-1}
\end{aligned}
\end{equation}

\begin{equation}
\begin{split}
&\operatorname{Im}\left\{ \hat{y}\cdot\mathbf{\overline{\overline{\mathbf{G}}}}\left(\mathbf{r_{0},r_{0},}\omega\right)\cdot\hat{y}\right\}=\left(\frac{1}{2\pi}\right)^{2} \intop_{0}^{k_{0}\sqrt{\mu\varepsilon_{>}}}dq_{\Vert}\intop_{0}^{2\pi}d\theta Re \Bigg\{  \\
&-\frac{q_{\Vert}\left(k_{0}^{2}\mu_{m}\varepsilon_{m}-q_{\Vert}^{2}cos^{2}\theta\right)}{\omega\mu_{0}\mu_{m}}\left(\frac{1}{2j\omega\varepsilon_{0}\varepsilon_{m}p_{m}}+B_{m}^{y}+A_{m}^{y}\right) \\
&+\left(A_{m}^{x}-B_{m}^{x}\right)\frac{jq_{\Vert}^{2}cos\theta p_{m}}{\omega\mu_{0}\mu_{m}}\Bigg\} .\label{eq:Gyy-1}
\end{split}
\end{equation}

\begin{figure}
\centering
\includegraphics[width=8.5cm]{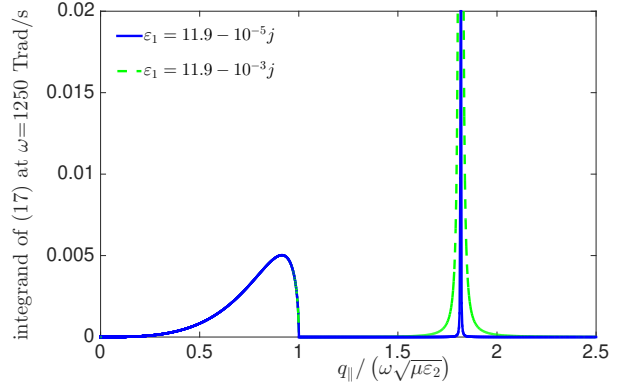}

\caption{\label{fig:integrand} The integrand of (\ref{eq:Gxx-1}) for a dielectric
slab with $\varepsilon_{1}=11.9$ immersed in $\varepsilon_{2}=2.1.$
The vertical source with $\omega=1250\,\mathrm{\frac{Trad}{s}}$ is
located at the center of the slab. y-axis is truncated and the maximum
intensities of the peaks are not shown. }
\end{figure}

As an example, consider a three layers structure consisting of a dielectric
slab with $\varepsilon_{1}=11.9$ and the thickness of $T_{1}=0.3\,\mu m$
immersed in a medium with $\varepsilon_{2}=2.1.$ For a vertical source
at the center of the slab, the integrand of (\ref{eq:Gxx-1}) is shown
in Fig. \ref{fig:integrand}. As expected, the wave components with
$q_{\Vert}<\omega\sqrt{\mu_{0}\varepsilon_{0}\varepsilon_{2}}$, which
can travel in both media without attenuation, are non-zero in the
integrand and contribute to the LDOS. Moreover, there is a single
mode supported at $q_{\Vert}\simeq0.75\,\omega\sqrt{\mu_{0}\varepsilon_{0}\varepsilon_{1}}$
which is the guided mode inside the slab and shows up as a singularity
in the integrand (a delta function). The analytical integrations of
such singularities is possible using complex plane analysis and Cauchy's
integration theorem. However, a simpler solution for our purpose is
to add a small loss to the medium. This moves the singularity(ies)
slightly away from the real axis in the complex $\left(\mathrm{Re}\left\{ q_{\Vert}\right\} ,\mathrm{Im}\left\{ q_{\Vert}\right\} \right)$
plane. As a result, the integrand of (\ref{eq:Gxx-1}) includes a
sharp peak at the singularity's $q_{\Vert}$. Adding more loss makes
the peak broader and decreases its maximum intensity. This is in favor
of the numerical calculations with the cost of becoming farther from
modeling a lossless structure. Figure \ref{fig:integrand} shows the
integrand for two loss values. The y- axis is truncated to clarify
the broadening of the peak by increasing the loss. The same argument
applies to the integrand of (\ref{eq:Gyy-1}). We continue with adding
$10^{-5}j$ to the permittivities throughout the rest of this paper.

Equations (\ref{eq:LDOS2}), (\ref{eq:Gxx-1}), and (\ref{eq:Gyy-1}),
give the LDOS inside a 1D layered structure for horizontal and vertical
sources. Since the calculations include two-fold integrations as well
as the inversion of a $4\left(N-1\right)\times4\left(N-1\right)$
matrix, its computational cost grows rapidly as more layers are added.
However, depending on the 1D PC parameters, a finite number of layers
(with manageable complexity) may suffice to model the LDOS of an infinite
1D PC. For example, in the example discussed in the next section,
it turns out we only need two layers on each side of the source region
in order to obtain accurate LDOS (and DOS) values in the desired frequency
range. 

\section*{example}

\begin{figure}
\centering
\includegraphics[width=6.5cm]{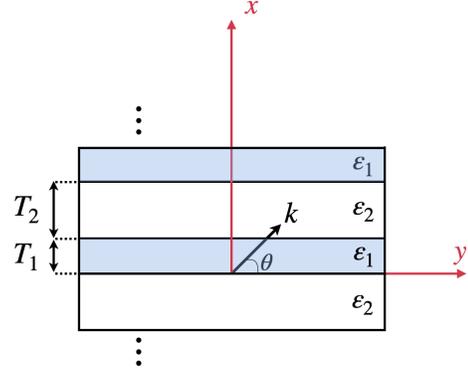}

\caption{\label{fig: 1DPC geometry }The 1D PC geometry. The y- and z- dimensions
are infinitely large. Parameters, in the example section, are $T_{1}=300\:nm,$
$T_{2}=700\:nm,$ $\varepsilon_{1}=11.9,$ $\varepsilon_{2}=2.1.$}
\end{figure}

Consider a 1D PC shown in Fig. \ref{fig: 1DPC geometry } with permittivities
of $\varepsilon_{2}=11.9$ (silicon), $\varepsilon_{2}=2.1$ (silicon
dioxide), and thicknesses of $T_{1}=300\,nm,$ $T_{2}=700\,nm$. Consider
a vertical (w.r.t. the layers boundaries) point source located at
the center of the $\varepsilon_{2}$ region. We approximate this geometry
with a finite number of layers around the source region similar to
Fig. \ref{fig:geometry}, and examine the LDOS by increasing the number
of layers.

\begin{figure}
\centering

\subfloat[]{ \includegraphics[width=8.5cm]{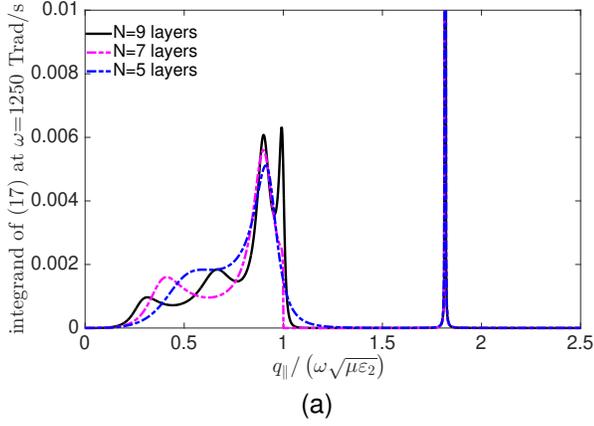}}
\vfil
\subfloat[]{ \includegraphics[width=4cm]{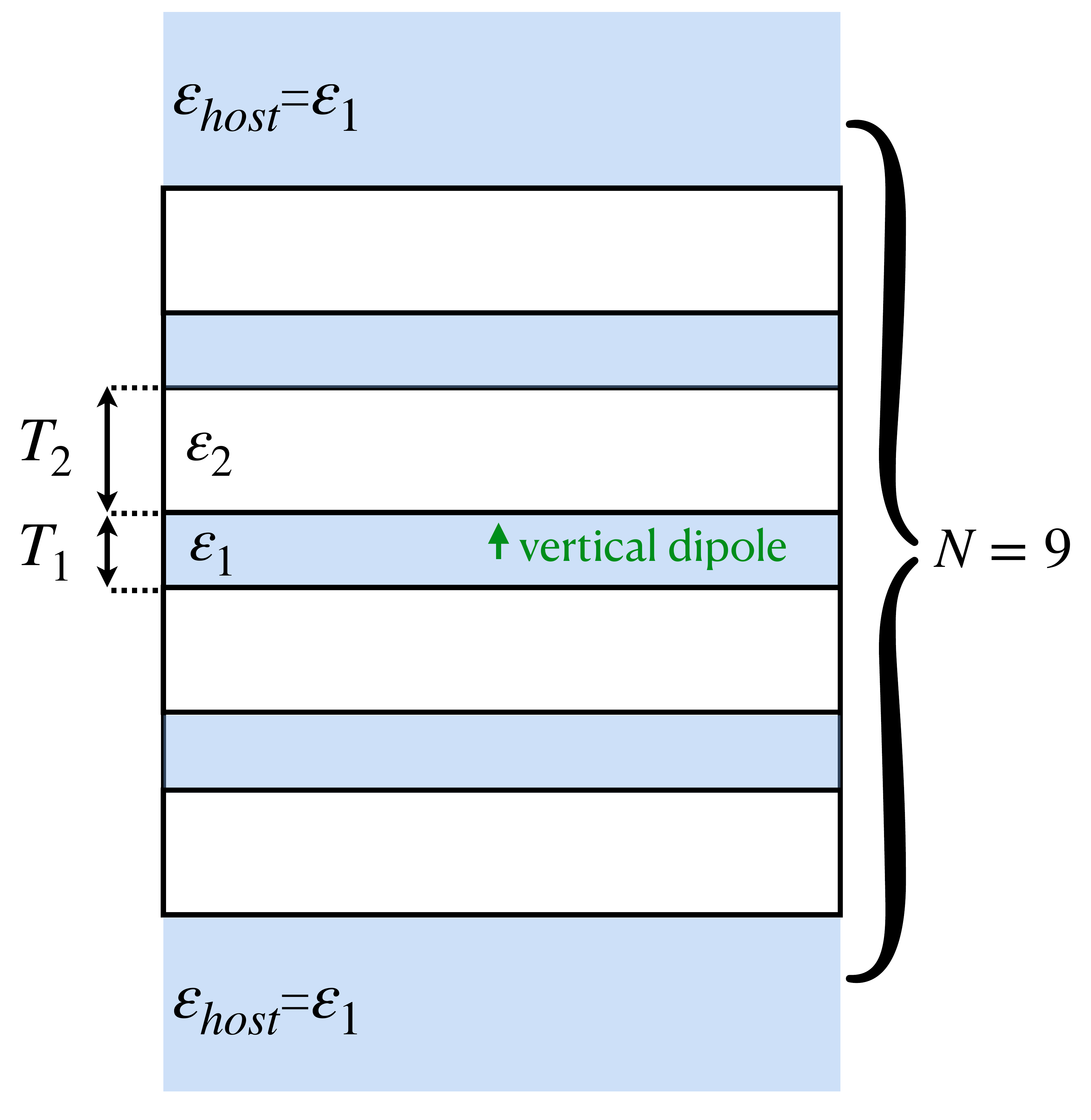}}

\caption{\label{fig:integrand_multilayer} a) the integrand of (\ref{eq:Gxx-1})
for a vertical source at the center of the $\varepsilon_{1}$ region
in Fig. \ref{fig: 1DPC geometry } geometry modeled with a finite
number of layers, (b) the nine layers model as a reference. Removing
one and two of the outermost layer from each side gives the seven
and five layers models, respectively. }
\end{figure}

Figure \ref{fig:integrand_multilayer} shows the integrand of \ref{eq:Gxx-1}
for a layered structure with five, seven, and nine layers, and parameters
similar to Fig. \ref{fig:geometry}. The vertical (x-directed) source
is at the center of the $\varepsilon_{1}$ region. The nine layers
geometry is also shown in Fig. \ref{fig:integrand_multilayer} as
a reference. By removing one and two of the outermost layers from
each side of this geometry, the seven and five layers geometries are
obtained, respectively. As Fig. \ref{fig:integrand_multilayer} shows,
the integrands of the three geometries have the sharp peak associated
with the wave-guided mode with the same $q_{\Vert}\simeq0.75\,\omega\sqrt{\mu_{0}\varepsilon_{0}\varepsilon_{1}}$
(and the same intensity, which is not shown in Fig. \ref{fig:integrand_multilayer}).
Other non-zero values of the integrand are associated with the propagating
modes which essentially become Bloch modes as the number of layers
increases sufficiently. 

\begin{figure}
\centering
\subfloat[]{ \includegraphics[width=8.5cm]{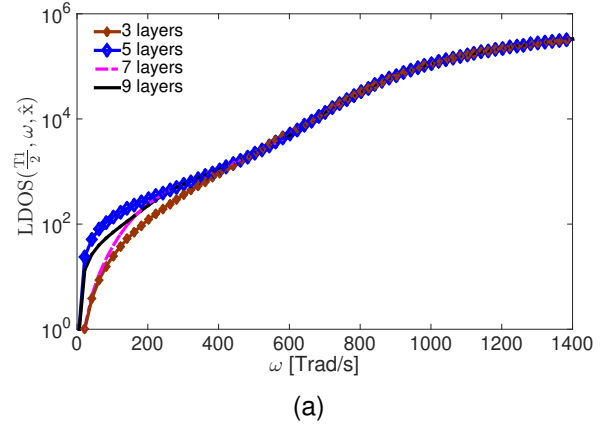}}
\vfill
\subfloat[]{ \includegraphics[width=8.5cm]{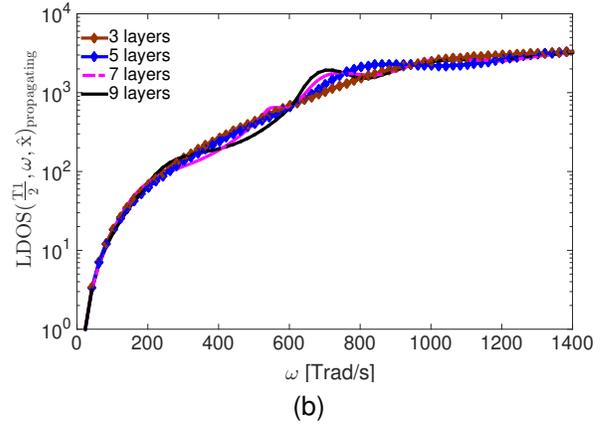}}
\caption{\label{fig:5and9layers}The LDOS (a) and $\mathrm{LDOS_{propagating}}$
(b) inside \ref{fig: 1DPC geometry } geometry modeled with different
number of layers. The source is vertical and at the center of the
middle $\varepsilon_{1}$ region.}
\end{figure}

Fig. \ref{fig:5and9layers} shows the calculated LDOS of the layered
structures using (\ref{eq:LDOS2}) and (\ref{eq:Gxx-1}). Fig. \ref{fig:5and9layers}
also includes the local density of propagating states ($\mathrm{LDOS_{propagating}}$)
by removing the wave-guided mode's contribution to the LDOS. This
quantity should merge to the local density of Bloch states, $\mathrm{LDOS_{Bloch}},$
as the number of layers grows sufficiently. It can be seen that the
LDOS of the five layers geometry does not change noticeably above
$\omega=250\,Trad/s$ by adding more layers. This is mostly because
the wave-guided mode's contribution dominates propagating modes' contribution
to the LDOS. As a result, we may use the five layers geometry to study
the LDOS and DOS of the infinite 1D PC shown in Fig. \ref{fig: 1DPC geometry }.
However, approximating Bloch mode densities of the 1D PC based on
the $\mathrm{LDOS_{propagating}}$ of the five layers geometry will
have a limited accuracy, as Fig. \ref{fig:integrand_multilayer} (b)
shows. We will discuss this issue, along with a solution later. The
same conclusions hold with a different source location and orientation.
In the following, we continue with the five layers model, and compare
the calculated LDOS and $\mathrm{LDOS_{propagating}}$ with alternative
methods. 

\begin{figure}
\centering
\subfloat[]{ \includegraphics[width=8.5cm]{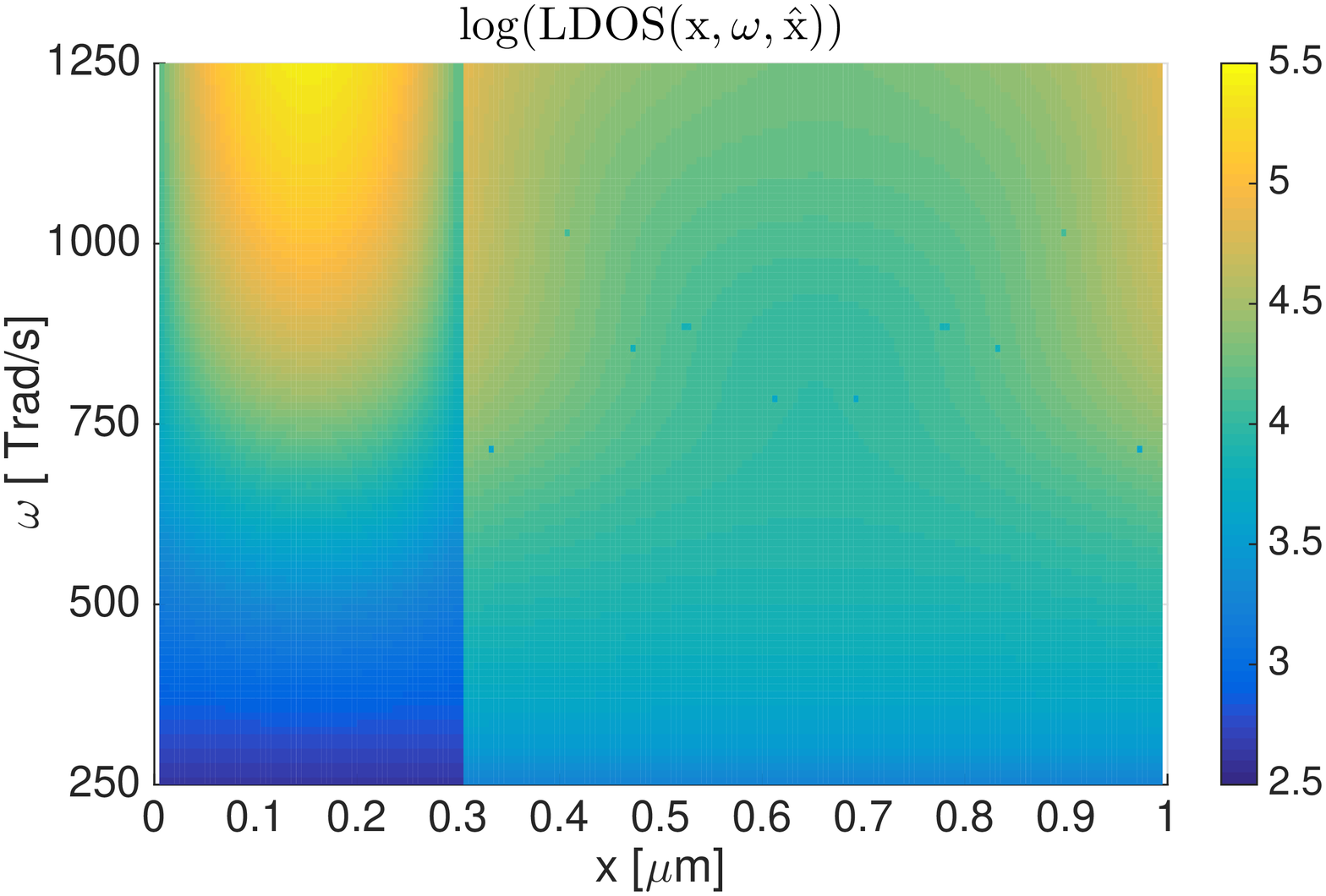}}
\vfill
\subfloat[]{ \includegraphics[width=8.5cm]{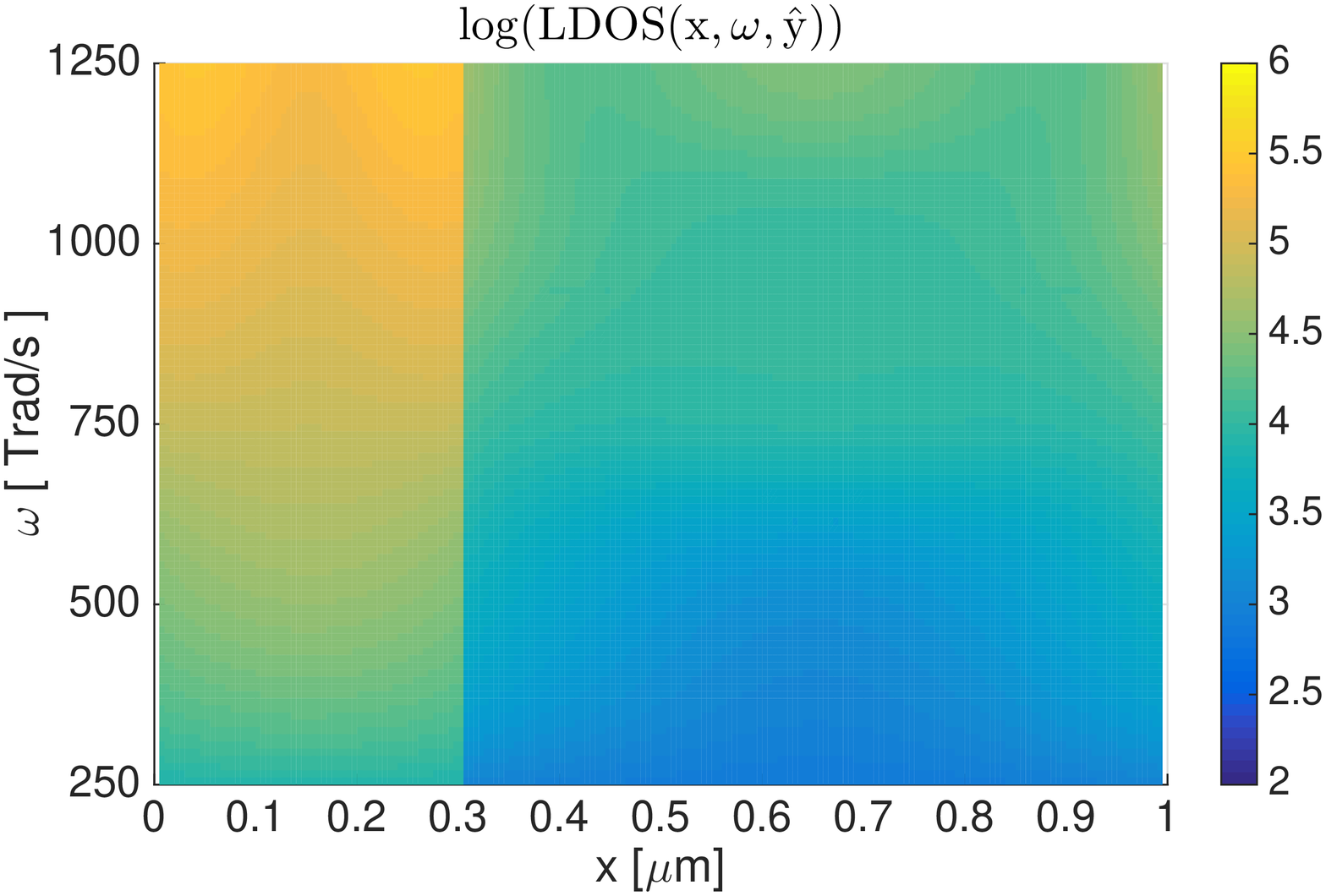}}
\vfill
\subfloat[]{ \includegraphics[width=8.5cm]{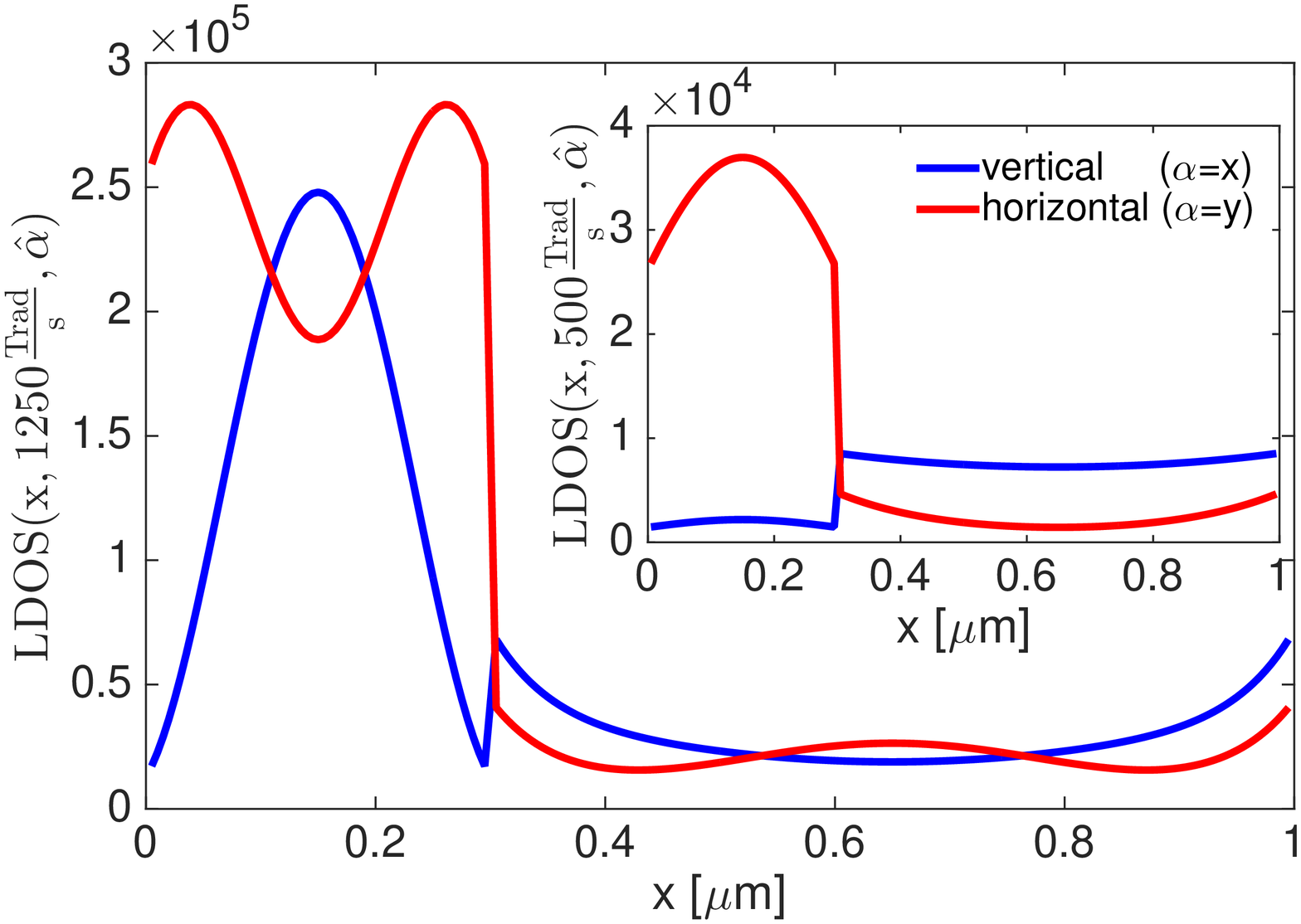}}

\caption{\label{fig:LDOS(200THz)} The LDOS of the 1D PC shown in Fig. \ref{fig: 1DPC geometry }
for a vertical source (a), a horizontal source (b), and at
$\omega=500,1250\,Trad/s$ (c). }
\end{figure}

Figure \ref{fig:LDOS(200THz)} shows the LDOS of the 1D PC for both
horizontal and vertical sources as a function of the source location
and radial frequency. Depending on the frequency, LDOS in the higher
permittivity region may become smaller than the lower permittivity
region. This will not be surprising if we consider that, beside the
number of available modes, the source-mode coupling intensity is involved
in the LDOS calculations. The larger LDOS at the edges of the $\varepsilon_{1}$
region at $\omega=1250\,Trad/s$ is consistent with the scalar calculations
in \cite{moroz1999minima}.

\begin{figure}
\centering
\subfloat[]{ \includegraphics[width=8.5cm]{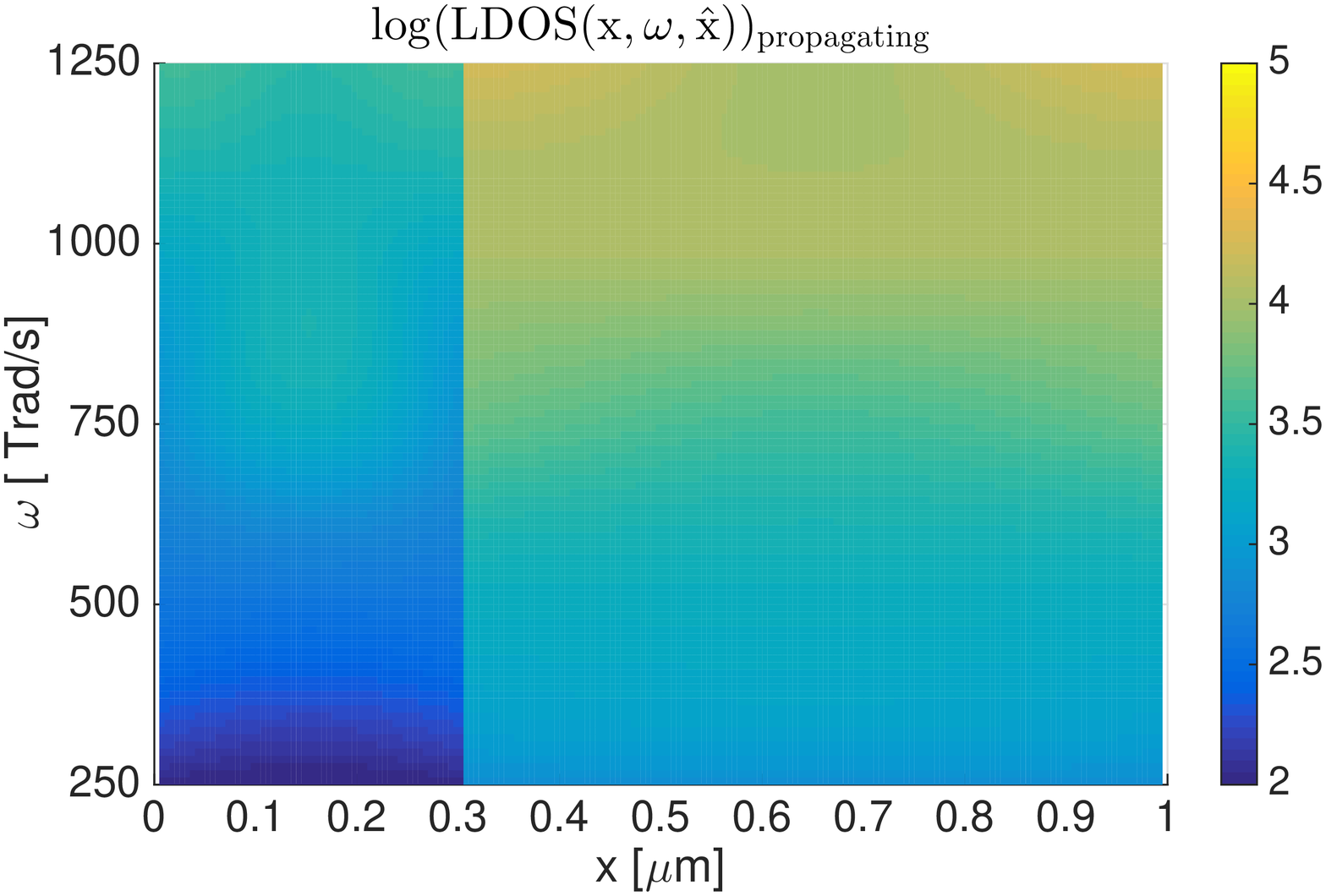}}
\vfill
\subfloat[]{ \includegraphics[width=8.5cm]{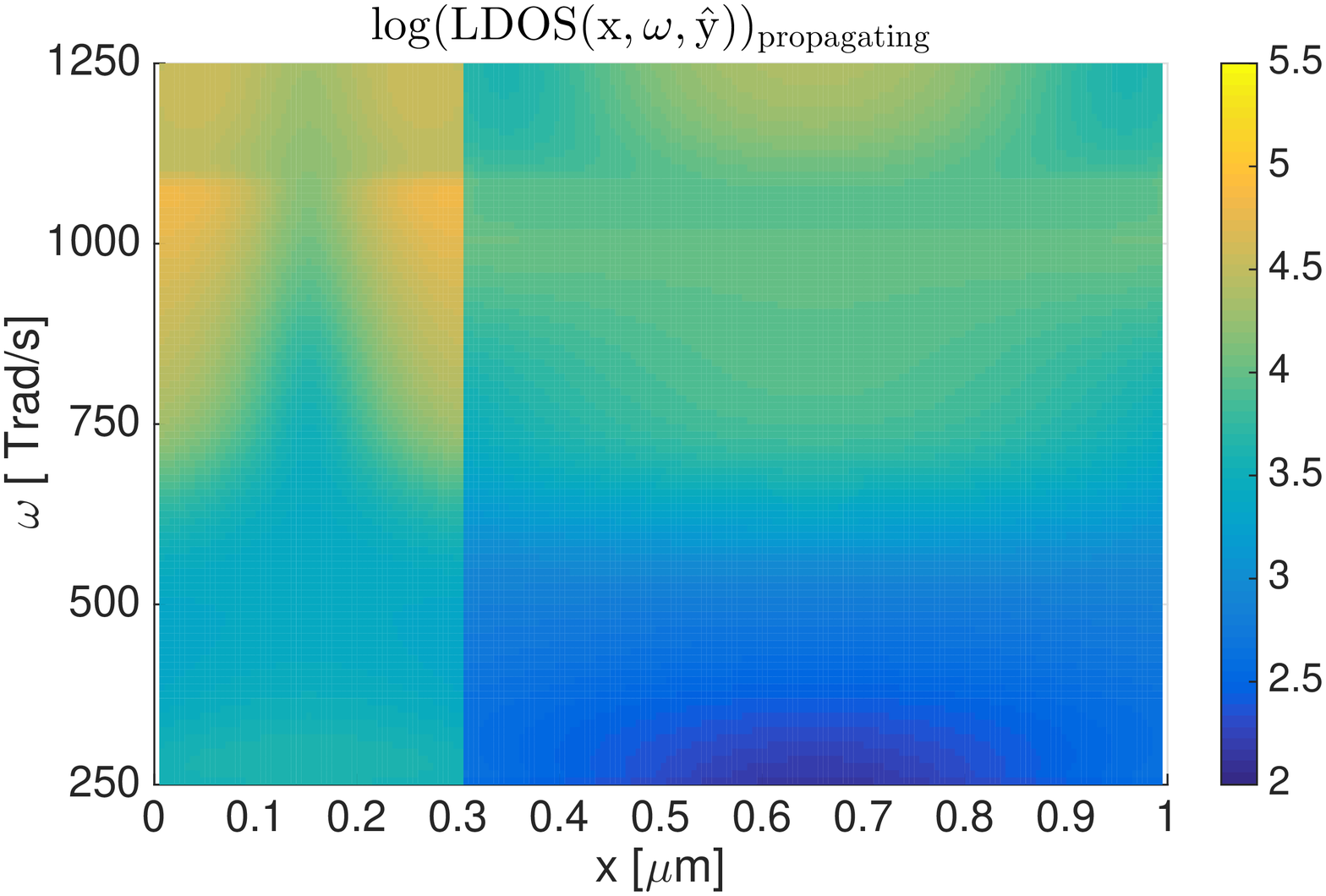}}
\vfill
\subfloat[]{ \includegraphics[width=8.5cm]{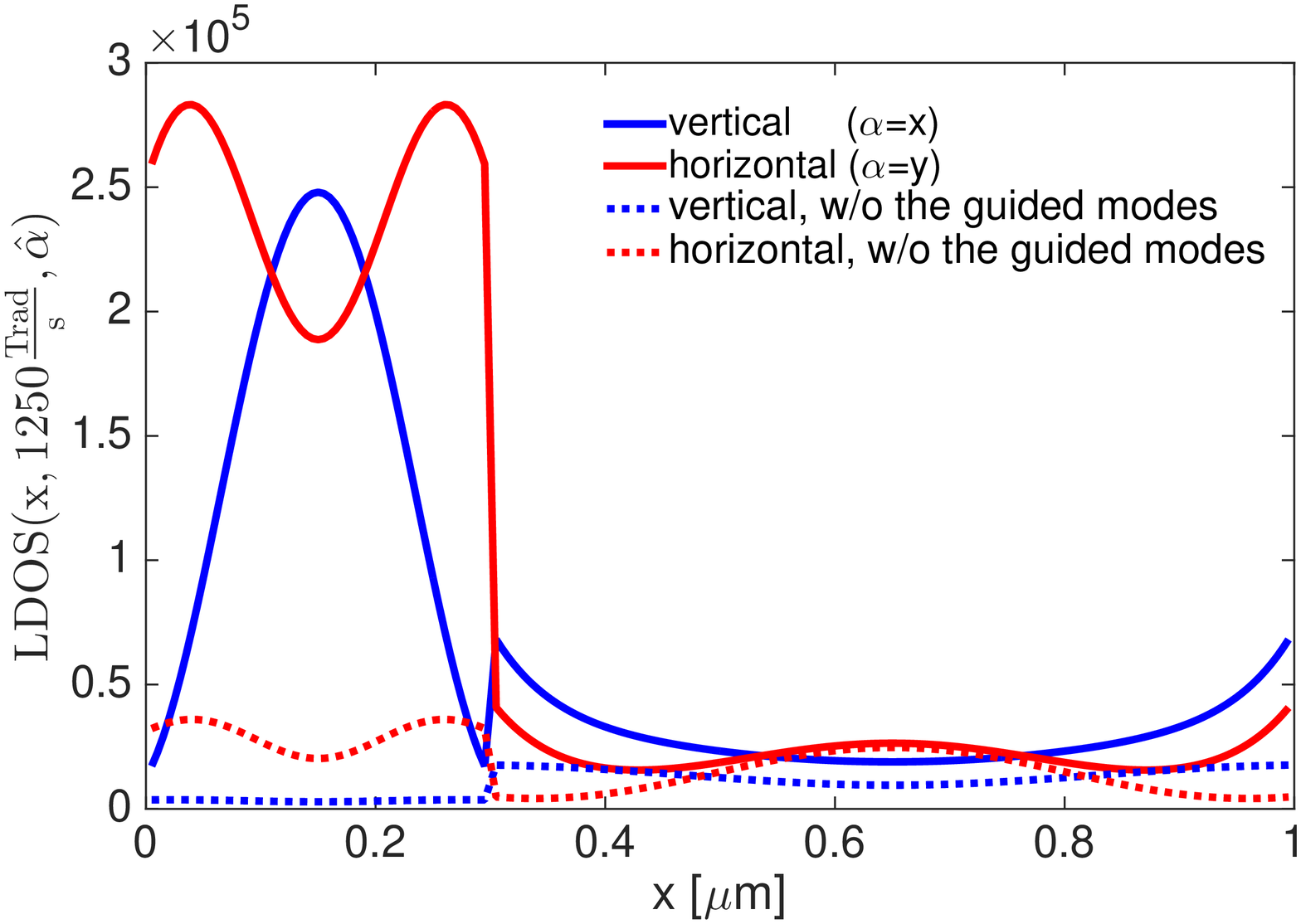}}

\caption{\label{fig:LDOS w/o TIR} The $\mathrm{LDOS_{propagating}}$ of the
five layer geometry for a vertical source (a) and a horizontal
source (b). (c) is the LDOS with and without the
guided modes at $\omega=1250\,Trad/S$. }
\end{figure}

As discussed earlier, the LDOS in a 1D PC is due to both Bloch modes
propagation carrying energy away from the source as well as the guided
modes confined mostly within the $\varepsilon_{2}$ region which do
not leave the structure (in case of a finite size in x- dimension)
\cite{wubs2002local}. The contribution of these guided modes to
the LDOS can be subtracted by removing the singularities of the Green's
tensor's integrand (we call this quantity $\mathrm{LDOS_{propagating}}$).
Figure \ref{fig:LDOS w/o TIR} shows the $\mathrm{LDOS_{propagating}}$
of the five layers geometry. Later we compare this quantity with the
Bloch density of the infinite 1D PC. Comparison between Fig. \ref{fig:LDOS(200THz)}
and Fig. \ref{fig:LDOS w/o TIR} confirms that the guided modes are
the dominant contributors to the LDOS, especially in the layers with
the higher permittivity. As a clarification, Fig. \ref{fig:LDOS w/o TIR}
also includes a comparison between the LDOSs with and without including
the guided modes at $\omega=1250\,Trad/s$. As expected, the guided
modes' contribution to the LDOS becomes minimized in the central locations
of the lower permittivity layer. 

\begin{figure}
\centering
\subfloat[]{ \includegraphics[width=8.5cm]{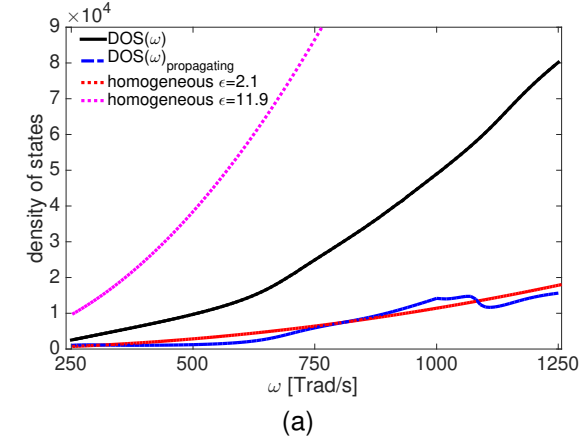}}
\vfill
\subfloat[]{ \includegraphics[width=8.5cm]{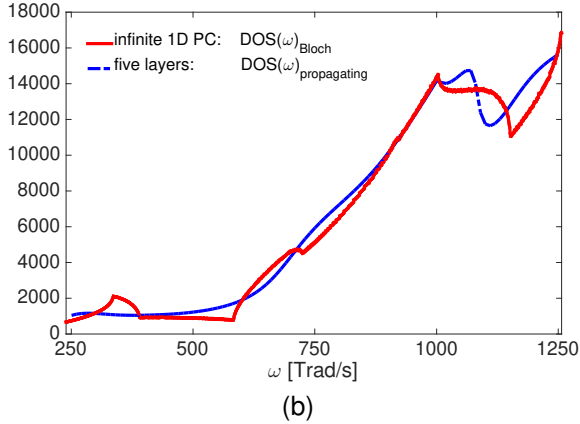}}
\caption{\label{fig:DOS} (a) the DOS and $\mathrm{DOS_{propagating}}$ of
the 1D PC shown in Fig. \ref{fig: 1DPC geometry } using the five
layers model, (b) comparison between $\mathrm{DOS_{propagating}}$
(of the five layers model) and $\mathrm{DOS_{Bloch}}$(of an infinite
1D PC). }
\end{figure}

Figure \ref{fig:DOS} (a) shows the calculated DOS and $\mathrm{DOS_{\mathrm{propagating}}}$
of the layered geometry using (\ref{eq:DOS-1}). It also includes
DOS of isotropic homogeneous media with permittivities of 2.1 and
11.9, as a reference. These DOSs are the estimated values for an infinite
1D PC based on the five layered structure calculations. As discussed
earlier, the LDOS and DOS calculations of the five layered structure
(at $\omega>250\,\mathrm{\frac{Trad}{s}}$) are very close to an infinite
1D PC with the same material properties (Fig. \ref{fig:5and9layers}(a)).
This was because these quantities are mainly determined by the guided
modes within the higher permittivity layers. However, as Fig. \ref{fig:5and9layers}(b)
showed, using $\mathrm{DOS_{\mathrm{propagating}}}$ of the five layers
structure to approximate $\mathrm{DOS_{\mathrm{Bloch}}}$ of the 1D
PC has limited accuracy and needed to be studied. To do so, we calculate
$\mathrm{DOS_{\mathrm{Bloch}}}$ of the infinite 1D PC using its dispersion
relations {[}Archived document{]}. A non-magnetic 1D PC as shown in
Fig. \ref{fig: 1DPC geometry }, has Bloch modes dispersion relations
as \cite{qi2017complex,joannopoulos1995photonic}

\begin{equation}
\begin{aligned}
cos\left(\left( T_{1}+T_{2} \right) k_{x} \right)&=cos\left(k_{x1}T_{1}\right)cos\left(k_{x2}T_{2}\right)\\
&-0.5\left(\frac{p_{2}}{p_{1}}+\frac{p_{1}}{p_{2}}\right)sin\left(k_{x2}d_{2}\right)sin\left(k_{x1}d_{1}\right),\label{eq:kx-1}
\end{aligned}
\end{equation}
where 

\begin{equation}
k_{xi}=\sqrt{k_{i}^{2}-k_{y}^{2}-k_{z}^{2}};\:p_{i}=\begin{cases}
\begin{array}{c}
\frac{k_{xi}}{\omega\mu_{0}}\\
\frac{\omega\varepsilon_{0}\varepsilon_{i}}{k_{xi}}
\end{array} & \begin{array}{c}
TE\\
TM
\end{array}\end{cases},
\end{equation}
and transverse electric (TE) and magnetic (TM) modes are defined with
respect to the y-z plane (interface plane). The Bloch density of states
can be calculated as 
\begin{equation}
\begin{aligned}
 \mathrm{DOS_{\mathrm{Bloch}}}\left(\omega\right)&= \sum_{j=TE,TM}\frac{\omega\varepsilon_{2}}{4c^{2}\pi^{2}}\intop_{0}^{\pi}d\theta\\
 &\left|\left(cos\theta\frac{\partial k_{x}^{j}\left(\omega,\theta\right)}{\partial\theta}+\omega sin\theta\frac{\partial k_{x}^{j}\left(\omega,\theta\right)}{\partial\omega}\right)cos\theta\right|.\label{eq:DOS_PC-1}
\end{aligned}
\end{equation}
where $\varepsilon_{2}<\varepsilon_{1}$. 

Figure \ref{fig:DOS}(b) shows the comparison between $\mathrm{DOS_{\mathrm{Bloch}}}$
of an infinite 1D PC (using (\ref{eq:DOS_PC-1})) and $\mathrm{DOS_{\mathrm{propagating}}}$
of the five layers geometry. Although the two quantities remain close
in the entire frequency interval, $\mathrm{DOS_{\mathrm{propagating}}}$
misses some of the derivative discontinuities in $\mathrm{DOS_{\mathrm{Bloch}}}$.
Figure \ref{fig:DOS}(b) also reveals the amount of inaccuracy
that presents in the total DOS calculation (black curve in Fig. \ref{fig:DOS}(a)).
In fact, we can remove such inaccuracy by replacing $\mathrm{DOS_{\mathrm{propagating}}}$
contribution in the total DOS calculation with $\mathrm{DOS_{\mathrm{Bloch}}}$.
This is not shown in figures, as the change was not noticeable (since
the DOS is dominated by the guided modes). 

\section*{FEM verification and comparing (\ref{eq:LDOS2}) with (\ref{eq:LDOS}) }

\begin{figure}
\centering
\subfloat[]{ \includegraphics[width=8.5cm]{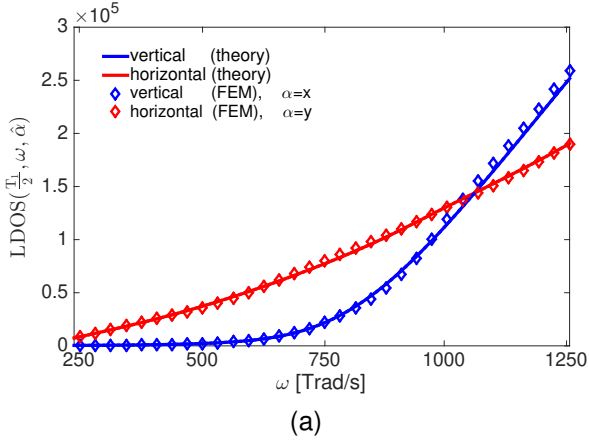}}
\vfill
\subfloat[]{ \includegraphics[width=8.5cm]{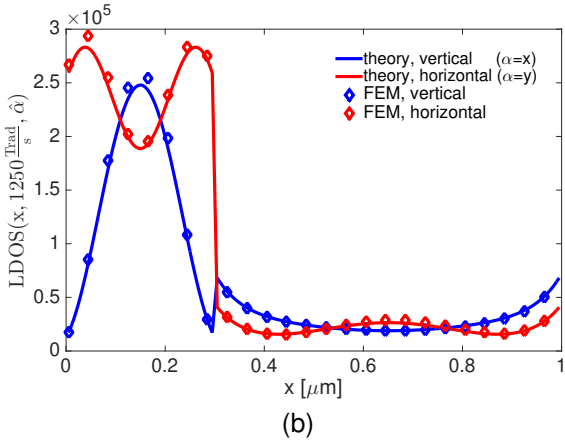}}

\caption{\label{fig:full-wave} Theoretical calculation of the LDOS versus
the full-wave solver results based on FEM at the center of the $\varepsilon_{1}$
region (a) and at $\omega=1250\,\mathrm{\frac{Trad}{s}}$ (b). }
\end{figure}

\begin{figure}
\centering
\subfloat[]{ \includegraphics[width=8.5cm]{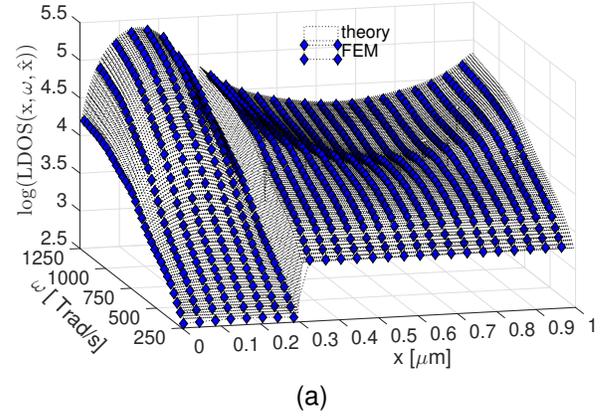}}
\vfill
\subfloat[]{ \includegraphics[width=8.5cm]{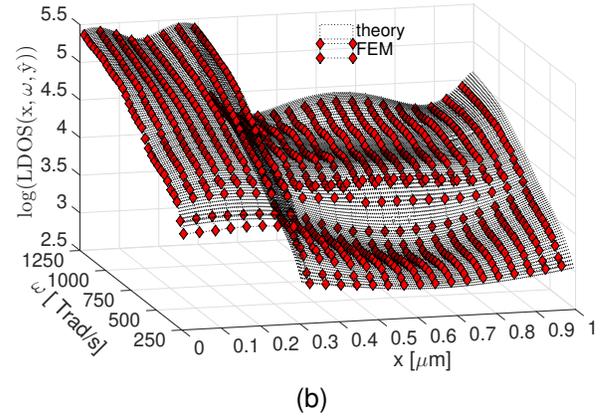}}
\caption{\label{fig:FEM_full} Theoretical calculation of the LDOS versus the
full-wave solver results based on FEM as a function of radial frequency
and location for a vertical (a) and horizontal emitter (b).}
\end{figure}

In order to verify the LDOS calculations, the 1D PC structure is numerically
solved using a full-wave solver based on FEM \cite{COMSOL} and the
results are compared in Figs. \ref{fig:full-wave} and \ref{fig:FEM_full}.
In the FEM modeler, a small cylinder with the surface area of $\delta s$
is place at the source location (we used hight and perimeter of $10\,\mathrm{nm}$)
with its axis aligned to $\hat{\alpha}$ direction. A surface electric/magnetic
current ($J_{\alpha}^{e}/J_{\alpha}^{m}$) is enforced on the surface
of the cylinder, directed along its axis. The real part of the electric/magnetic
field component co-directed with the applied current is averaged over
the cylinder volume and is extracted ($\mathrm{Re}\left\{ E_{\alpha}^{ave}\right\} /\mathrm{Re}\left\{ H_{\alpha}^{ave}\right\} $).
The necessary components of the Green's tensors for the LDOS calculations
are simply 

\begin{equation}
\begin{aligned}
&\left|\omega\mu_{0}\mu_{r}\operatorname{Im}\left(\hat{\alpha}\cdot\mathbf{\overline{\overline{\mathbf{G}}}_{E}}\left(\mathbf{r}_{0},\mathbf{r_{0}},\omega\right)\cdot\hat{\alpha}\right)\right|=\left|\frac{\mathrm{Re}\left\{ E_{\alpha}^{ave}\right\} }{J_{\alpha}^{e}\delta s}\right| \\
&\left|\omega\varepsilon_{0}\varepsilon_{r}\operatorname{Im}\left(\hat{\alpha}\cdot\mathbf{\overline{\overline{\mathbf{G}}}_{H}}\left(\mathbf{r}_{0},\mathbf{r_{0}},\omega\right)\cdot\hat{\alpha}\right)\right|=\left|\frac{\mathrm{Re}\left\{ H_{\alpha}^{ave}\right\} }{J_{\alpha}^{m}\delta s}\right|.
\end{aligned}
\end{equation}
 The lateral dimensions of the structure is truncated to $1.25\lambda$,
and perfect matched layer (PML) are used as the boundary conditions.
Increasing the number of layers beyond five, or the lateral dimensions
of the geometry does not change the calculated LDOSs noticeably, which
ensures the convergence of the calculations. Figures \ref{fig:full-wave}
and \ref{fig:FEM_full} show good agreement between the theoretical
calculations and the FEM results. 

\begin{figure}
\centering
\subfloat[]{ \includegraphics[width=8.5cm]{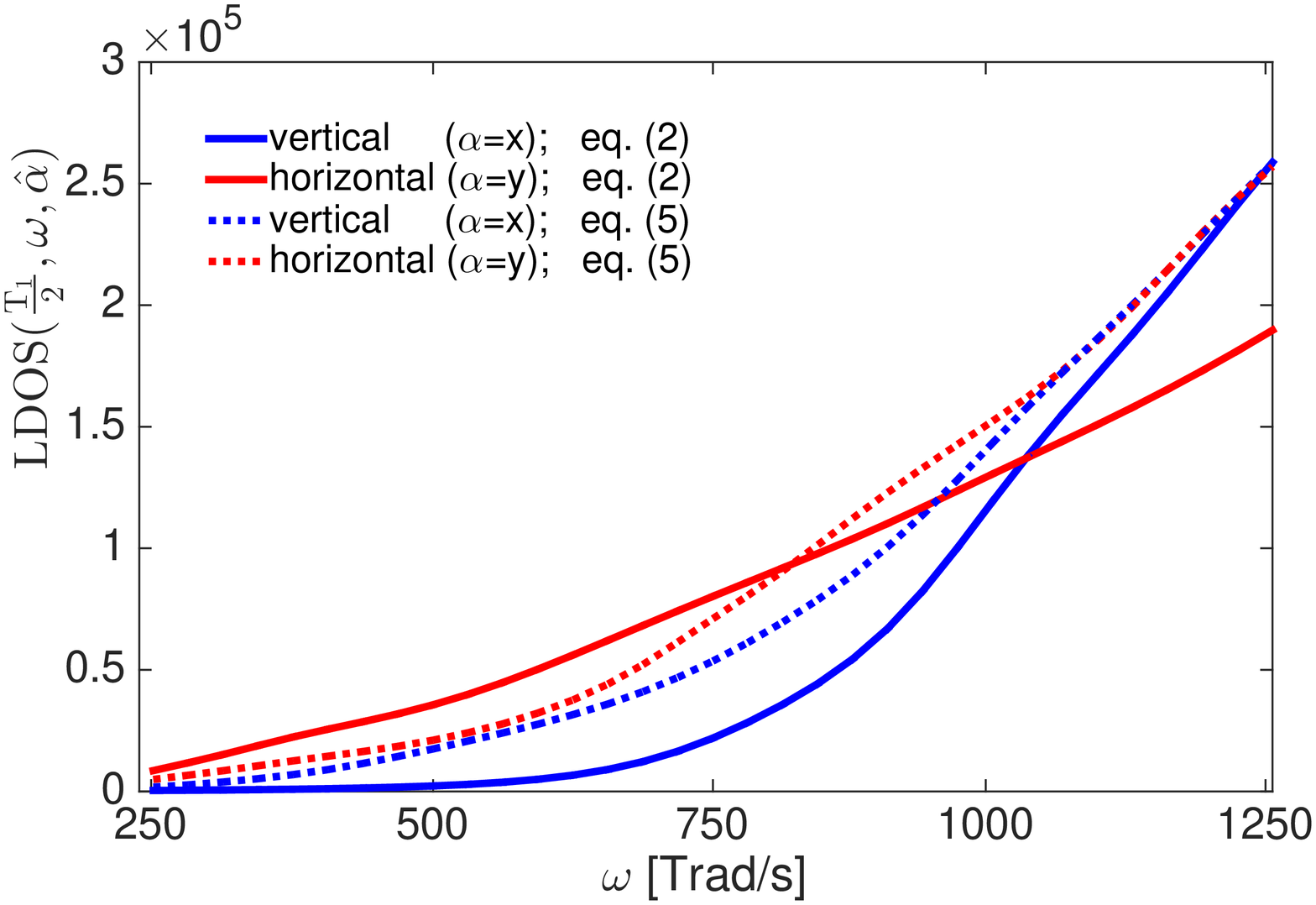}}
\vfill
\subfloat[]{ \includegraphics[width=8.5cm]{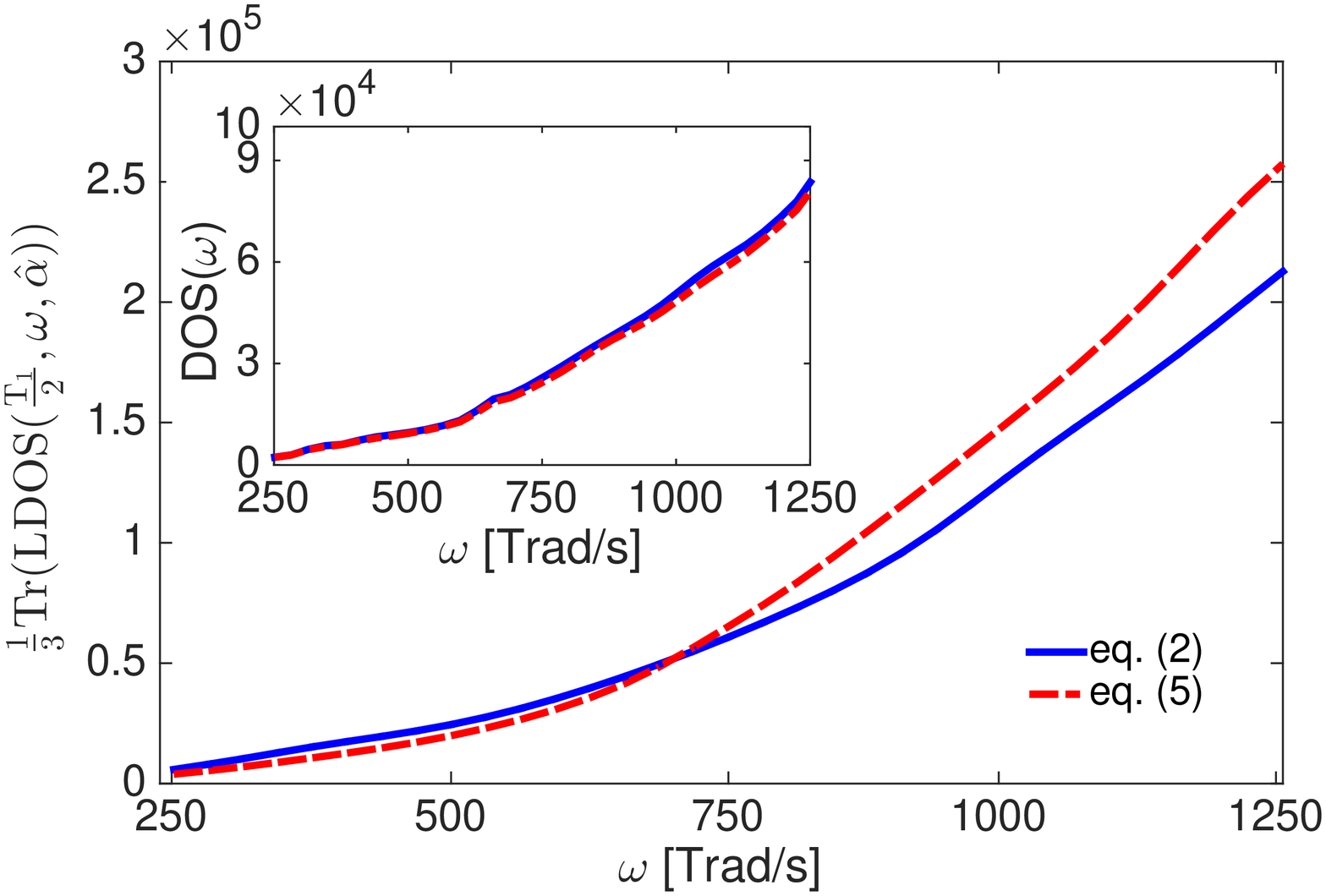}}

\caption{\label{fig:comparison} (a) the LDOS of the 1D PC at the center
of the $\varepsilon_{2}$ region using (\ref{eq:LDOS2}) and (\ref{eq:LDOS}).
(b) the average LDOS in all directions at the center of the $\varepsilon_{1}$
region, and the DOS (insert) using (\ref{eq:LDOS2}) and (\ref{eq:LDOS}). }

\end{figure}

As mentioned earlier, the LDOS calculations in this paper were based
on the electric point source only, i.e. (\ref{eq:LDOS2}), with the
implicit assumption that the contributions of the electric and magnetic
point sources to the LDOS are equal. As discussed in details in \cite{joulain2003definition},
in a general medium (e.g. anisotropic), (\ref{eq:LDOS2}) is an approximation,
and (\ref{eq:LDOS}) is the accurate LDOS expression. As a comparison,
Fig. \ref{fig:comparison}(a) shows the LDOS calculations based on
the electric and magnetic point sources (vertical and horizontal orientations)
at the center of the $\varepsilon_{1}$ region. It is clear that the
1D PC LDOS calculations based on (\ref{eq:LDOS2}) and (\ref{eq:LDOS})
have differences as the electric and magnetic dipoles' couplings to
the available modes, at a specific location, are generally different.
However, as Fig. \ref{fig:comparison} shows, the average LDOS in
all directions is less dependent on the considered point sources.
The calculated DOS based on (\ref{eq:LDOS2}) is almost as accurate
as using (\ref{eq:LDOS}). Accurate calculation of the LDOS based
on (\ref{eq:LDOS}), although follows a similar procedure, is beyond
the scope of this document. 

To briefly discuss the application of the obtained results in the
context of thermal radiation, consider a 1D PC as the emitting body.
The DOS for the emitters deep inside the 1D PC, as apparent from Fig.
\ref{fig:DOS}, is not exactly proportional to $\omega^{2}.$ However,
the $\mathrm{DOS_{Bloch}}$ is the more relevant quantity in thermal
studies rather than the total DOS since the photons in the guided
modes cannot leave the 1D PC. The $\mathrm{DOS_{Bloch}}$ in Fig.
\ref{fig:DOS} proves that emission from a 1D PC can have a non-Planckian
distribution (note that Planckian distribution is a result of $\omega^{2}$
dependence of the DOS). However, the obtained results are not conclusive
about whether the emission can surpass the Planck's limit because of
the reflection at the surface of a finite-size 1D PC. In order to
find the surface reflection impact, we can find the electric (far)
field in the outer most layer (i.e. layer N in Fig. \ref{fig:geometry})
generated by a point source inside the layered structure using (\ref{eq:Hertzian dipole})
to (\ref{eq:H}). The ratio of the emitted power into the outer half-space
to the total power departing the point source is the quantity which
can approximate the surface reflection impact, and determines whether
the radiations to the outer space can exceed Planck's limit. This is
a subject for a follow up study. 

\section*{Conclusion}

The formulations were provided for the exact LDOS calculations inside
a layered structure based on its Green's tensor. The relations were
used to calculate LDOS, DOS, and $\mathrm{DOS_{Bloch}}$ of a 1D PC.
Using parameters of a practical 1D PC at visible range, the quantities
were calculated for a five layer structure. It was shown that LDOS
and DOS of the 1D PC can be well-approximated by the five layers geometry
above a certain frequency. The exact $\mathrm{DOS_{Bloch}}$ inside
an infinite 1D PC was also calculated based on its dispersion relations,
and was compared with the approximation by the five layers geometry.
The results were also verified with a wave-solver based on FEM. The
calculated $\mathrm{DOS_{Bloch}}$ inside the 1D PC verifies the non-Planckian
thermal emission inside a 1D PC. 

\bibliographystyle{IEEEtran}
\bibliography{Referencesx_Thermal}

\end{document}